\documentclass[aps,reprint,superscriptaddress,article]{revtex4-2}

\usepackage[colorlinks,linkcolor=blue,anchorcolor=blue,citecolor=blue,urlcolor=blue]{hyperref}
\usepackage{graphicx}
\usepackage{epstopdf}
\usepackage{dcolumn}
\usepackage{multirow}
\usepackage{diagbox}
\usepackage{array}
\usepackage{bm}
\usepackage[title]{appendix}
\makeatletter

\newcommand{\Rmnum}[1]{\expandafter\@slowromancap\romannumeral #1@}
\makeatother

\begin{document}

\preprint{APS/123-QED}

\title{Anomalous Phonon in Charge-Density-Wave Phase of Kagome Metal CsV$_{3}$Sb$_{5}$}

\author{Han-Yu Wang}
\affiliation{\it Key Laboratory of Materials Physics,Institute of Solid State Physics, HFIPS, Chinese Academy of Sciences,Hefei 230031, China}
\affiliation{\it Science Island Branch of Graduate School, University of Science and Technology of China, Hefei 230026, China}

\author{Xiao-Cheng Bai}
\affiliation{\it Key Laboratory of Materials Physics,Institute of Solid State Physics, HFIPS, Chinese Academy of Sciences,Hefei 230031, China}
\affiliation{\it Science Island Branch of Graduate School, University of Science and Technology of China, Hefei 230026, China}

\author{Wen-Feng Wu}
\affiliation{\it Key Laboratory of Materials Physics,Institute of Solid State Physics, HFIPS, Chinese Academy of Sciences,Hefei 230031, China}
\affiliation{\it Science Island Branch of Graduate School, University of Science and Technology of China, Hefei 230026, China}

\author{Zhi Zeng}
\affiliation{\it Key Laboratory of Materials Physics,Institute of Solid State Physics, HFIPS, Chinese Academy of Sciences,Hefei 230031, China}
\affiliation{\it Science Island Branch of Graduate School, University of Science and Technology of China, Hefei 230026, China}

\author{Da-Yong Liu}
\affiliation{\it Department of Physics, School of Sciences, Nantong University, Nantong 226019, China}

\author{Liang-Jian Zou}
\email{zou@theory.issp.ac.cn}
\affiliation{\it Key Laboratory of Materials Physics,Institute of Solid State Physics, HFIPS, Chinese Academy of Sciences,Hefei 230031, China}
\affiliation{\it Science Island Branch of Graduate School, University of Science and Technology of China, Hefei 230026, China}


\begin{abstract}

  CsV$_3$Sb$_5$, a notable compound within the kagome family, is renowned for its topological and superconducting properties, as well as its detection of local magnetic field and anomalous Hall effect in experiments. However, the origin of this local magnetic field is still veiled. In this study, we employ the first-principles calculations to investigate the atomic vibration in both the pristine and the charge-density-wave phases of CsV$_3$Sb$_5$. Our analysis reveals the presence of ``anomalous phonons" in these structures, these phonon induce the circular vibration of atoms, contributing to the phonon magnetic moments and subsequently to the observed the local magnetic fields. Additionally, we observe that lattice distortion in the charge-density-wave phase amplifies these circular vibrations, resulting in a stronger local magnetic field, particularly from the vanadium atoms. This investigation not only reveals the potential relation between lattice distortion and atomic polarization but also offers a novel idea to understand the origin of local magnetic moment in CsV$_3$Sb$_5$.

\end{abstract}
%
\maketitle

\emph{Introduction.-}
The kagome lattice, which consists of planar triangles and hexagons structures, was first introduced to statistical physics by Syozi \cite{Syozi1951-PTP}.
In past decades, many unusual properties and behaviours, such as geometrical frustration \cite{Ramirez1994-ARMS}, quantum spin liquid states \cite{Yan2011-Science}, charge-density-wave (CDW) \cite{van2021-NatMater,Gao2009-PRB}, topological insulator \cite{Gao2009-PRB}, and unconventional superconductivity (SC) \cite{Ko2009-PRB}, {\it etc.}, were proposed for this lattice, suggesting kagome compounds may display numerous quantum phases.
Very recently, some new kagome metals AV$_{3}$Sb$_{5}$ (A=K, Rb, and Cs) have been discovered \cite{Ortiz2019-PRM}. Universal chiral CDW order coexisting with unconventional SC at low temperatures was observed \cite{Ortiz2020-PRL,Chen2021-Nature,Chen2021-PRL} in these compounds.
The CDW phases show $2\times2$ charge order, and the compounds display three SC phases with increasing hydrostatic pressure; these kagome metals demonstrate $Z_{2}$ topological surface state was observed \cite{Yong2022-SciBull}, and so on. These compounds integrate many important concepts in condensed matter physics, raising great attention and debates on the microscopic origins of the CDW order and SC order.

Among these unusual properties, two phenomena are especially attractive. On the one hand, both the neutron scattering experiment \cite{Xie2022-PRB} and the muon spin relaxation ($\mu$SR) experiment \cite{Kenney2021-JPCM,Li2021-arXiv} demonstrated the absence of measurable local spin moments or magnetic correlations, implying a perfect nonmagnetic phase in kagome metals AV$_{3}$Sb$_{5}$.
However, a giant anomalous Hall effect (AHE) was observed, concomitant with the charge order at low temperatures \cite{Mielke2022-Nature,Yu2021-PRB,Shumiya2021-PRB}, indicating the presence of intrinsic magnetic field. On the other hand, the time-reversal symmetry breaking (TRSB) also occurs with the chiral CDW ordering  in the scanning tunneling microscopy experiment \cite{Shumiya2021-PRB,Jiang2021-NatMater,Zhao2021-Nature}. Whereafter, chiral orbital flux or loop current was put forward by Feng {\it et al.} \cite{Feng2021-SciBull} and Denner {\it et al.} \cite{Denner2021-PRL} for addressing the CDW ordering and TRSB; on the contrary, several experiments denied the presence of the chiral flux current in bulk AV$_{3}$Sb$_{5}$ \cite{Li2022-NatPhys,Li2022-PRB}.
The topical issue thus arises what causes the AHE and the TRSB in the CDW phase of these nonmagnetic metals.

It is widely recognized that when charged ions undergo circularly or elliptically vibrations, they generate extra orbital angular momenta and extra orbital magnetic moments. Recently, Zhang {\it et al.} proposed that coherent motions of two components of charged ions could result in chirality of phonon \cite{Zhang2015-PRL}. This specialized phonon, known as chiral phonon, emerge from the global spatial inversion or the time-reversal symmetry breaking within specific system \cite{Zhang2015-PRL,Zhu2018-Science}, including hexagonal lattice \cite{Zhang2015-PRL,Ptok2021-PRB,Wang2021-JPCM} and kagome lattice \cite{Chen2019-PRB}.
The circular or elliptical vibrations of chiral phonons thus may contribute extra orbital angular momenta and orbital magnetic moment, leading to the phonon magnetic moment \cite{Jiaming2023-Science,Juraschek2019-PRM,Xiong2022-PRB}.
Conversely, in certain conventional materials preserving the spatial inversion and time-reversal symmetries, charged ions perhaps still exhibit circular or elliptical vibration in particular phonons. However, due to the presence of opposite polarizations of atomic vibrations, the total polarizations of these phonons are zero, thus they do not have chirality \cite{Zhang2015-PRL,Chen2019-PRB,Ptok2021-PRB,Wang2021-JPCM}.
To distinguish this kind of phonon from the conventional one, here, we term it ``anomalous phonon", which has local orbital magnetic moment like chiral phonon. In this context, the relationship between the chiral phonon and the anomalous one resembles that between ferromagnet and antiferromagnet.

In this {\it Letter}, we investigate the atomic polarization in the pristine and the CDW phases of CsV$_{3}$Sb$_{5}$, utilizing an extension from chiral phonon to anomalous phonon for topologically kagome metals. We reveal the existence of anomalous phonons in these systems and show the vanadium atoms contribute extra phonon magnetic moment in the CDW phase. Such an orbital magnetic moment provide a novel idea to explain the emergence of measurable local magnetic field \cite{Kenney2021-JPCM,Li2021-arXiv}, AHE \cite{Mielke2022-Nature,Yu2021-PRB,Shumiya2021-PRB} and TRSB \cite{Shumiya2021-PRB,Jiang2021-NatMater,Zhao2021-Nature} in low-T phase of AV$_{3}$Sb$_{5}$.

%

\begin{figure}[htbp]
\includegraphics[width=8.5cm]{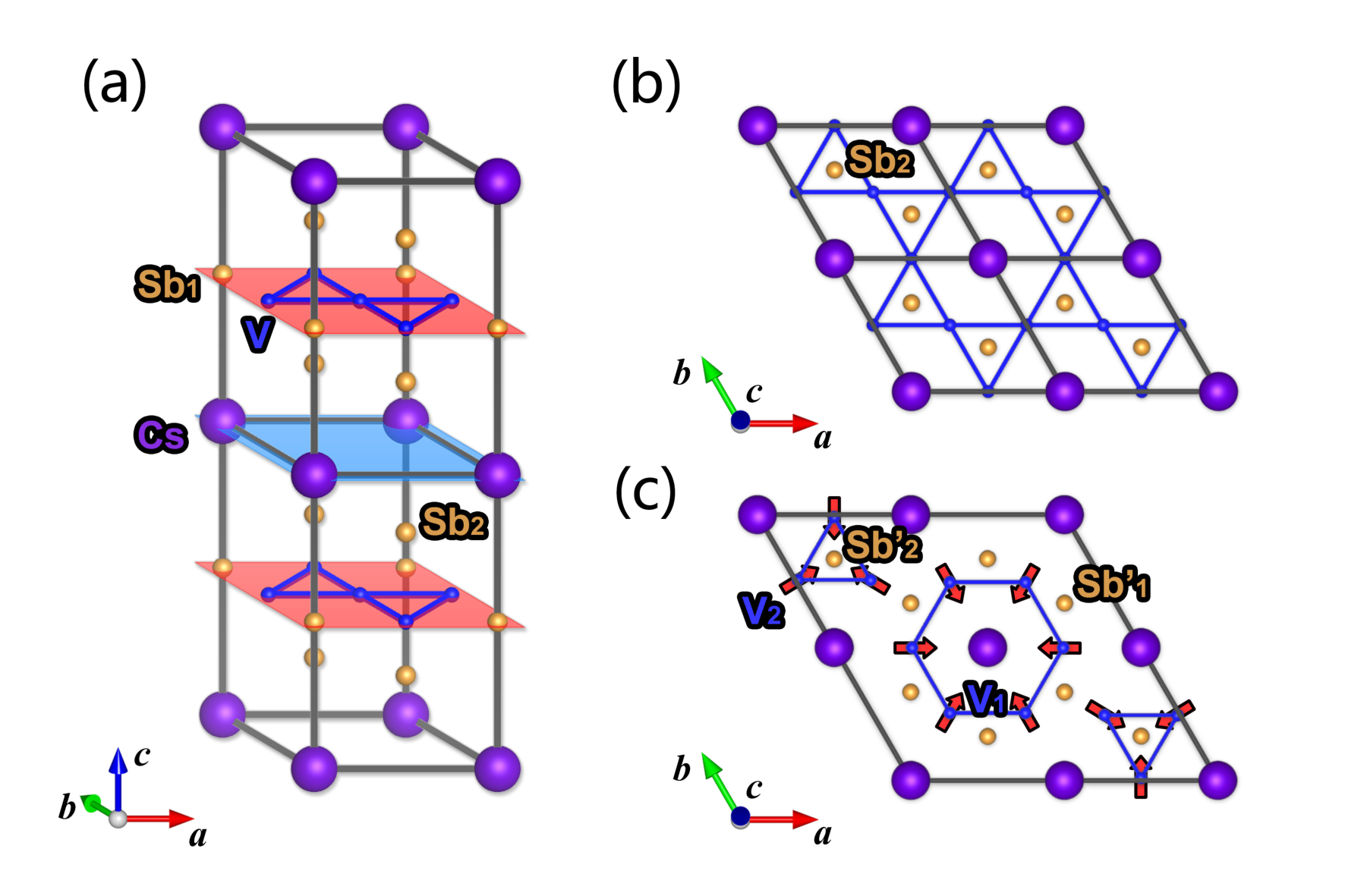}
\caption{\label{P-1}
  Crystal structures of CsV$_{3}$Sb$_{5}$. (a) The oblique view of the pristine phase. The red and blue layers indicate the kagome plane and Cs plane, respectively. (b) The top view of the pristine phase with the $2\times2$ unit cells. (c) The top view of the ISD $2\times2$ CDW phase. The red arrows mark the distortion direction of V atoms in $ab$-plane.}
\end{figure}

\emph{Crystal structures.-}
To investigate the phonon of CsV$_{3}$Sb$_{5}$, we initially optimize the structures by the first-principles calculations.
Fig.\ref{P-1} (a) and (b) show the optimized pristine phase of CsV$_{3}$Sb$_{5}$. The structure showcases a layered arrangement of V-Sb atomic sheets interspersed with Cs atomic layers, and the V atoms construct the kagome plane in $ab$-plane. The crystal structure is classified to the $P6/mmm$ space group with $C_{6}$ symmetry. Cs and V atoms each have an equivalent site, while Sb atoms have two inequivalent sites, which are labeled as Sb$_{1}$ and Sb$_{2}$ in Fig.\ref{P-1} (a) and (b).
The lattice constants of the $ab$-plane and the $c$-axis are 5.45 \AA ($a$,$b$) and 9.30 \AA ($c$), respectively, in agreement with experimental observation at high temperature \cite{Zhao2021-Nature}.

As the temperature decreases to the critical value, CsV$_{3}$Sb$_{5}$ transits from high-temperature homogenous pristine phase to the low-temperature CDW phase. In this CDW phase, the V atoms undergo a shrinkage resembling an inverse Star of David (ISD) phase with the $2\times2$ supercell \cite{Tan2021-PRL,Ortiz2020-PRL}, as illustrated in Fig.\ref{P-1} (c).
Upon transitioning to the ISD $2\times2$ CDW phase, the six V atoms in hexagonal configuration move toward to the center, while the three V atoms in the triangle configuration shift outward, corresponds to a breathing-in phonon within the kagome plane. This optimized crystal structure of the CDW phase agrees with the prior findings \cite{Tan2021-PRL,Ortiz2020-PRL}, maintaining the $P6/mmm$ space group with $C_{6}$ symmetry. However, due to such lattice distortion, there emerge two inequivalent sites (V$_{1}$ and V$_{2}$) for the V atom, as well as two inequivalent sites (Sb'$_{1}$ and Sb'$_{2}$) for Sb$_{2}$ atom, which are labeled in Fig.\ref{P-1} (c).



\emph{Anomalous phonon.-}
Chen {\it et al.} \cite{Chen2019-PRB} reported the presence of chiral and anomalous phonons in the kagome lattice.
To ascertain whether anomalous phonons also exist in CsV$_{3}$Sb$_{5}$, we investigate the phonon dispersion and atomic polarization $S^{z}_{\alpha}$ of both the pristine and CDW phases within the first Brillouin zone (BZ), the magnitude of $S^{z}_{\alpha}$ indicates the extent of atomic circular vibration. For detailed calculations and discussions on the polarization $S^{z}_{\alpha}$, please refer to the {\it Supplemental Materials} Sec.S3.

\begin{figure}[htbp]
\includegraphics[width=8.5cm]{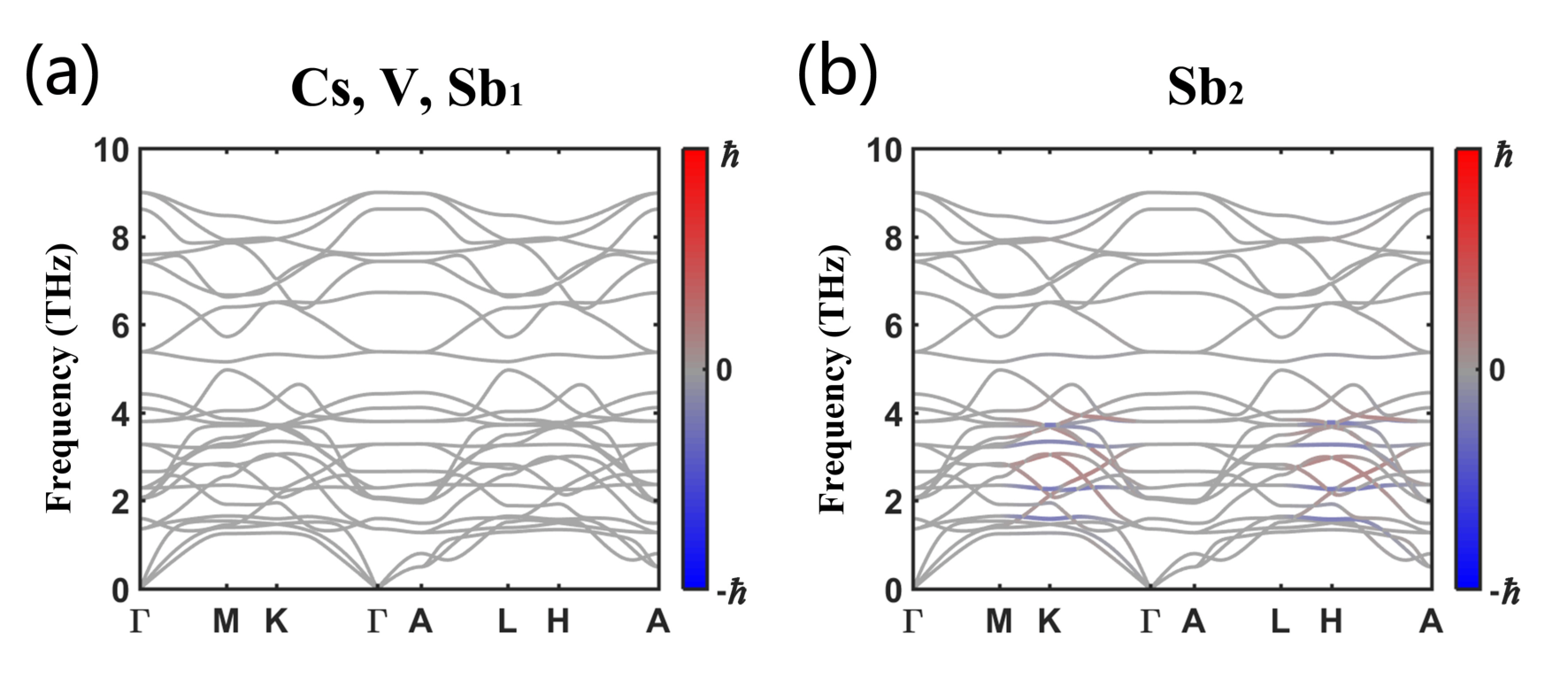}
\caption{\label{P-2}
   Phonon dispersions of the pristine phase with atomic polarization $S^{z}_{\alpha}$ along the high-symmetry path. (a) Phonon dispersion with the atomic polarization of Cs, V and Sb$_{1}$ atoms. (b) Phonon dispersion with the atomic polarization of Sb$_{2}$ atom. The red and blue parts represent the opposite direction of circular vibration.}
\end{figure}

Fig.\ref{P-2} shows the phonon dispersions with atomic polarization $S^{z}_{\alpha}$ distribution of the pristine phase in the high-symmetry path of the BZ. The unit cell comprises eight atoms, resulting in 24 phonon branches in the dispersion.
As shown in Fig.\ref{P-2} (a), all Cs, V and Sb$_{1}$ atoms exhibit zero polarization $S^{z}_{\alpha}$, indicating the absence of circular vibration for these atoms.
Conversely, Fig.\ref{P-2} (b) illustrates the polarization of an Sb$_{2}$ atom, this atom displays nonzero polarization $S^{z}_{\alpha}$ in certain phonons, signifying its engagement in circular vibration and the presence of anomalous phonons in the pristine phase. These anomalous phonons predominantly occur at low frequencies ranging from 1 THz to 4 THz.
It is noteworthy that Cs, V and Sb$_{1}$ atoms reside within the symmetric $ab$-plane, representing spatial inversion symmetry points. Therefore, the local atomic circular vibrations in these phonons may originate from the inherent asymmetry of these sites.

\begin{figure}[htbp]
\includegraphics[width=8.5cm]{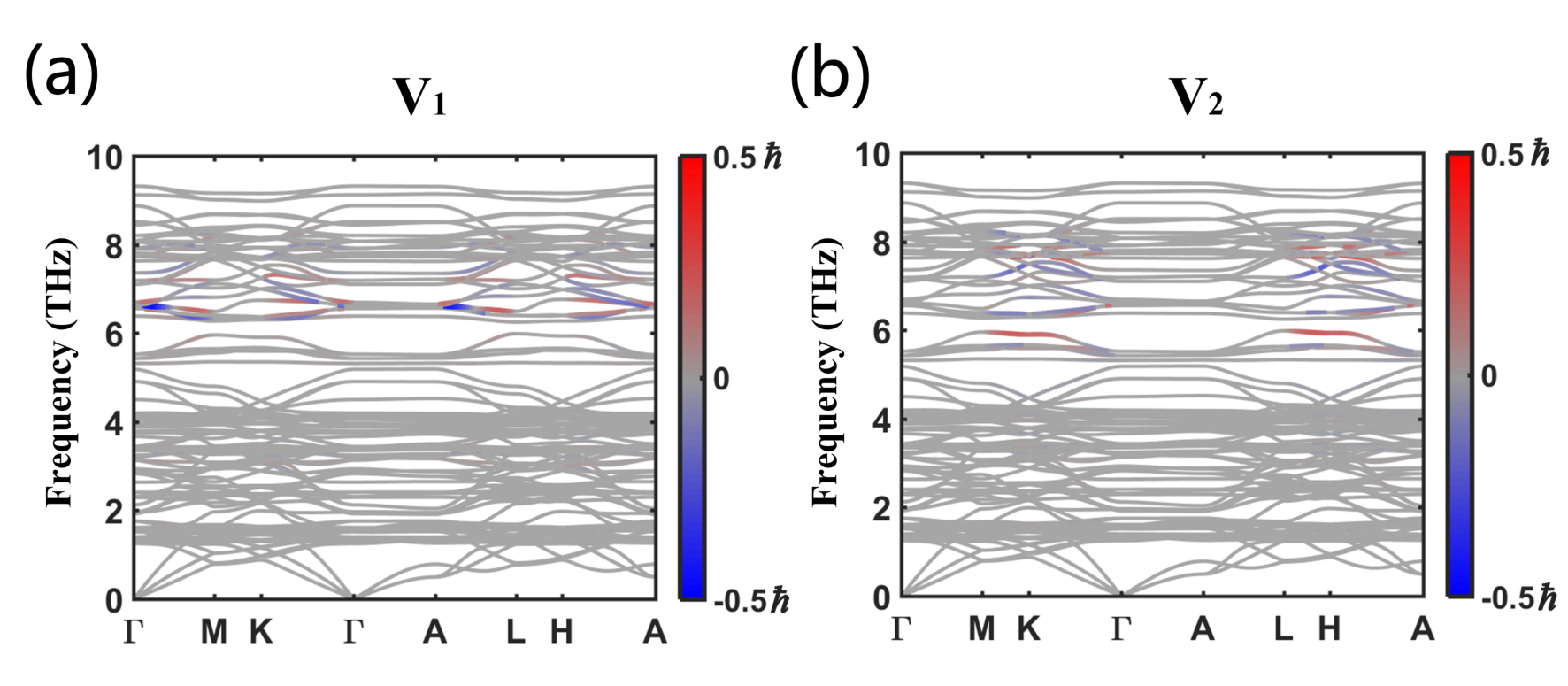}
\caption{\label{P-3}
   Phonon dispersions of the CDW phase with atomic polarization $S^{z}_{\alpha}$ along the high-symmetry path. (a) Phonon dispersion with the atomic polarization of V$_{1}$ atom. (b) Phonon dispersion with the atomic polarization of V$_{2}$ atom. The red and blue parts represent the opposite direction of circular vibration.}
\end{figure}

In the CDW phase, the Cs atoms occupy two inequivalent sites and the atomic polarizations of these atoms remain zero.
Similarly, the sites of Sb$_{1}$ atoms, which were equivalent in the pristine phase, evolute into two inequivalent sites in the CDW phase, with both exhibiting zero atomic polarization.
Moreover, the sites of Sb$_{2}$ atoms in the pristine phase also evolute to two inequivalent sites (Sb'$_{1}$ and Sb'$_{2}$) in the CDW phase, and the nonzero polarization distribution resembles that of the pristine phase.
Distinctively, the two inequivalent V$_{1}$ and V$_{2}$ atoms also display circular vibration in certain phonon modes. Fig.\ref{P-3} (a) and (b) show the phonon dispersions and atomic polarizations of V atoms in the CDW phase. Due to the distortion in the 2$\times$2 supercell, this dispersion comprises 96 phonon branches, with those involving circular vibrations of V atoms exhibiting higher frequencies compared to those involving circular vibrations of Sb atoms.
Compared to the pristine phase, V atoms experience the $ab$-plane distortion in the CDW phase. Consequently, the emergence of V atomic polarization stems from the distortion of the kagome plane.

These nonzero atomic polarizations indicate the presence of anomalous phonons in both the pristine and CDW phases of CsV$_{3}$Sb$_{5}$. However, according to our analysis, the total atomic polarizations of these anomalous phonon is zero, thus these anomalous phonons are not chiral phonons. We will delve into this discussion in detail in the subsequent section.

\emph{Atomic polarization and circular vibration.-}
Atomic polarization serves as a measure of the magnitude of circular vibration. To investigate the polarization values and the nature of atomic vibration, we numerically label these phonon modes based on their frequency magnitude ranging from low to high and focus on the high-symmetry $K$ point as an example, to elucidate the anomalous phonons.
Table.\ref{T1} lists the nonzero polarizations of atoms in select anomalous phonons.
Notably, Mode 8 in the pristine phase and Mode 27 in the CDW phase exhibit similar frequencies, suggesting an evolutionary relationship between them.
Besides, Sb'${_2}$ atoms evolve from Sb${_2}$ atoms following structural distortion. In these phonon modes, both Sb${_2}$ and Sb'${_2}$ atoms exhibit have nonzero atomic polarization.
Considering that the phonon eigenvector is normalized, it is appropriate to compare polarizations after expanding the unit cell of the pristine phase to a $2\times2$ supercell. Please refer to {\it Supplemental Materials} Sec.S3 for discussions on this matter. Following this expansion, the polarization value of each atom will be 1/4, i.e., 0.0625, which is half of the polarization value of Sb'$_{2}$ atoms in the CDW phase.
This indicates that the Sb'${_2}$ atoms in the CDW phase exhibit a greater degree of circular polarization compared to their counterparts.
Regarding the V$_{2}$ atoms in the CDW phase, although their polarization is relatively smaller compared to the Sb'$_{2}$ atoms, they still exhibit significant polarization properties, surpassing the polarization of Sb$_{2}$ atoms in the pristine phase. The corresponding energy of this anomalous phonon is higher, approximately 5.90 THz.

\begin{table*}[htbp]
  \setlength{\belowcaptionskip}{15pt}
  \caption{\label{T1} The magnitude of atomic polarization in specific anomalous phonon modes at the high-symmetry $K$ point.
  The sites of Sb$_{2}$, Sb'$_{2}$, and V$_{2}$ atoms are labeled in Fig.\ref{P-1} and \ref{P-4}.}
  \begin{ruledtabular}
  \begin{tabular}{m{5cm}<{\centering}|cc|ccc}
  Phase                  & \multicolumn{2}{m{4cm}<{\centering}|}{Pristine Phase} & \multicolumn{3}{m{6cm}<{\centering}}{CDW Phase}                                          \\ \hline
  Atom                  & \multicolumn{2}{c|}{Sb$_{2}$}                         & \multicolumn{2}{m{4cm}<{\centering}|}{Sb'$_{2}$}                              & V$_{2}$  \\ \hline
  Mode                   & \multicolumn{1}{m{2cm}<{\centering}|}{8}   & 16       & \multicolumn{1}{m{2cm}<{\centering}|}{27}        & \multicolumn{1}{c|}{61}    & 77       \\
  Frequency (THz)        & \multicolumn{1}{c|}{2.27}                  & 3.72     & \multicolumn{1}{c|}{2.35}                        & \multicolumn{1}{c|}{3.98}  & 5.90     \\
  Polarization ($\hbar$) & \multicolumn{1}{c|}{0.250}                 & 0.250    & \multicolumn{1}{c|}{0.127}                       & \multicolumn{1}{c|}{0.134} & 0.086
  \end{tabular}
  \end{ruledtabular}
\end{table*}

\begin{figure}[htbp]
\includegraphics[width=8.5cm]{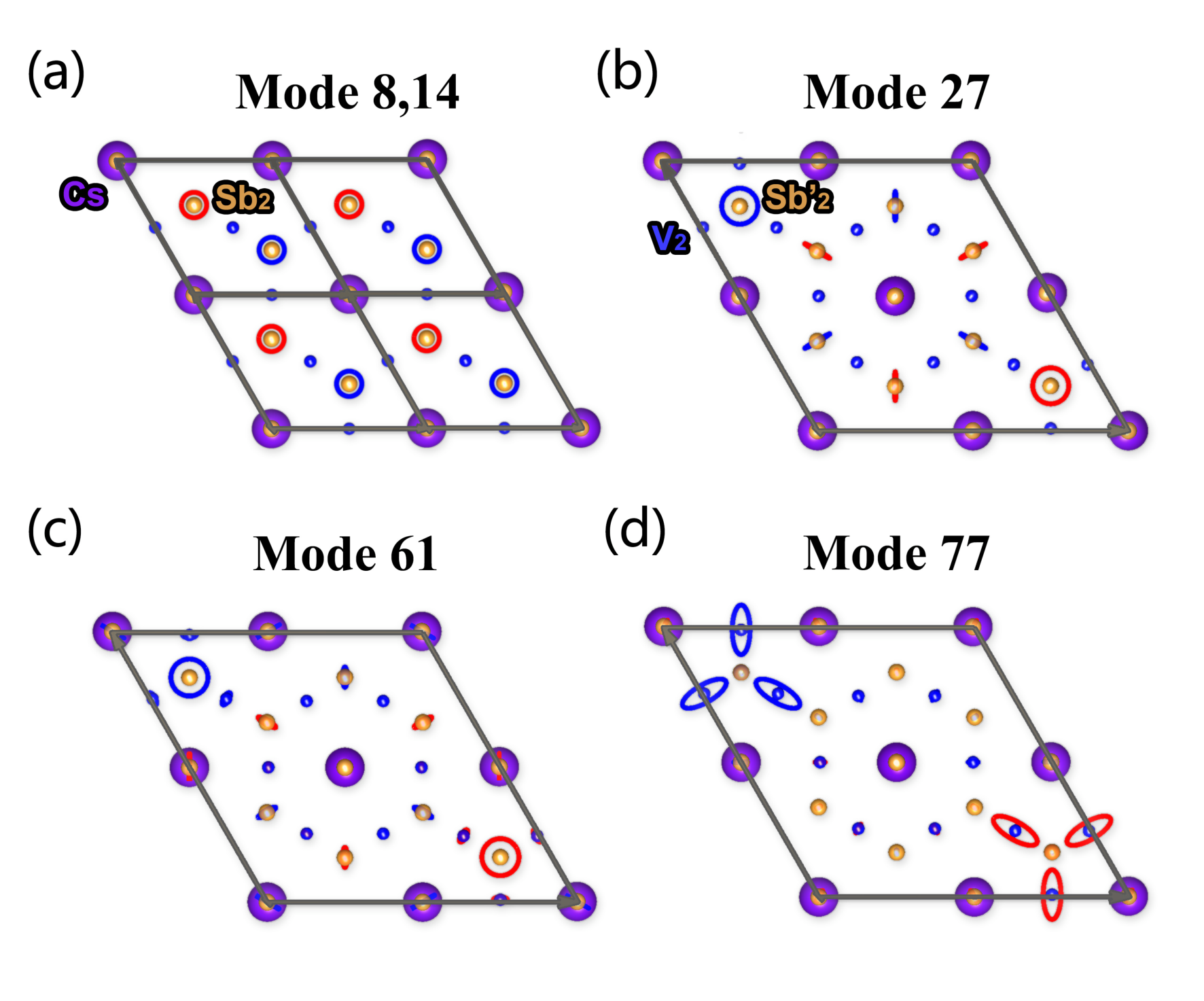}
\caption{\label{P-4}
   Atomic vibrations in specific anomalous phonons at the high-symmetry $K$ point. (a) The atomic vibrations of Mode 8 and 14 in the pristine phase. (b), (c), (d) The atomic vibration of Mode 27, 61 and 77 in the CDW phase, respectively. The radius of the circular trajectory around atoms are proportional to the amplitude of atomic circular vibration. The red and blue parts represent the opposite directions of atomic vibration.}
\end{figure}

To facilitate a more intuitive discussion, we depict the vibration trajectory of each atom in these phonon modes, as shown in Fig.\ref{P-4}.
The trajectories of these atoms maintain $C_{6}$ symmetry within these crystal structures, consistent with the symmetry of the crystal lattice.
In these phonon modes, for Sb${_2}$ and Sb'${_2}$ atoms, the vibration trajectories form complete circles within the $ab$-plane. Additionally, the circularly polarized Sb atoms are paired in a unit cell, with their vibration directions opposing each other. Thus, the total phonon polarization is zero, and there is no chirality present in this phonon mode, owing to the spatial inversion symmetry inherent in the system.
However, as discussed above, for the other Sb atoms located at spatial inversion symmetry points, they do not exhibit circular vibration.
For V atoms, their atomic vibrations are shown in Fig.\ref{P-4} (d).
In this phonon mode, except for V$_{2}$ atoms, the other atoms remain relatively stationary. This indicates that these V atoms predominantly contribute to the total energy of harmonic vibration in this mode.
Additionally, three adjacent V atoms vibrate in opposite directions to the other three adjacent V atoms, resulting in a total phonon polarization of zero.
Comparing the vibrations of Sb atoms in Fig.\ref{P-4} (a-c), V atoms exhibit a larger amplitude in one direction, although the polarization trajectory forms an ellipse. Thus, as shown in Table.\ref{T1}, its polarization value of V atom are smaller than those of Sb atoms in Fig.\ref{P-4} (c).
Moreover, we find that lager structural distortions induce greater polarization of V$_{2}$ atoms. For a detailed discussion, please refer to {\it Supplemental Materials} Sec.S5.

\emph{Local magnetic field.-}
Circular atomic vibration can result in phonon magnetic moment, which, typically, shares a similar order of magnitude with the nuclear magnetic moment \cite{Juraschek2019-PRM}.
Moreover, the $\mu$SR experiments have detected local magnetic fields in AV${3}$Sb${5}$, with magnitudes also comparable to the nuclear magnetic moment \cite{Kenney2021-JPCM}.
In this section, we explore the potential attribution of the observed local magnetic field in experiments to the phonon magnetic moment.

\begin{figure}[htbp]
\includegraphics[width=8.5cm]{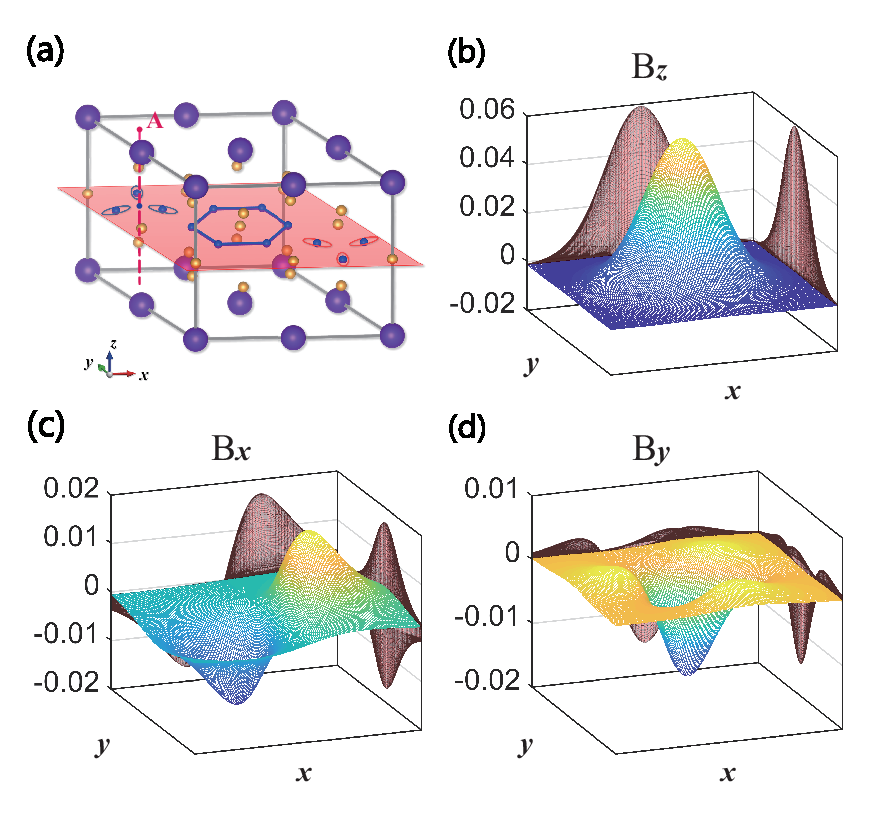}
\caption{\label{P-5}
  The simulation of local magnetic field distribution by vibrations of three neighbor V$_{2}$ atoms at the Cs plane in Mode 77. (a) Vibrations of atoms in V$_{2}$ site in Mode 77. The point {\bf A} is the projection of the center of the triangle constructed by three V atoms along $z$-axis in Cs plane. (b),(c),(d) The intensity distribution of the local magnetic field along the $z$, $x$ and $y$ directions around {\bf A} point, respectively.}
\end{figure}

The magnitude of the phonon moment correlate to the gyromagnetic ratio $\gamma_{\alpha}=e\mathbf{Z}_{\alpha}/(2M_{\alpha})$. However, Born effective charges $\mathbf{Z}_{\alpha}$ are ill-definition in metal, like CsV$_{3}$Sb$_{5}$ \cite{Gonze1997-PRB}.
Nonetheless, Born effective charges reflect the amount of carried charges of ions, and estimate them by calculating the charge density distribution. In the {\it Supplemental Materials} Sec.S4, we provide support for the charge density distribution of CsV$_{3}$Sb$_{5}$.
Compared to Sb atoms, V atoms have a smaller effective mass $M_{\alpha}$. Moreover, based on the charge density distribution, V atoms harbor numerous localized electrons, resulting in a larger magnitude of Born effective charges $\mathbf{Z}_{\alpha}$.
Thus, V atom has a greater gyromagnetic ratio $\gamma_{\alpha}$, leading a larger local phonon magnetic moment in circular vibration.

Yu {\it et al.} \cite{Li2021-arXiv} reported the detection of local magnetic fields along different directions in the Cs plane following the CDW phase transition, as observed through the $\mu$SR experiments. Here, we take Mode 77 as an example and analyze the local magnetic fields generated by the polarizations of V atoms in the Cs plane.
Figure.\ref{P-5} (a) shows the vibrations of V atoms in this mode. As mentioned in the previous section,  the vibration trajectory of V atoms forms an ellipse located in the $ab$-plane. To investigate the local field distribution generated by the three neighboring V atoms in a triangular configuration, we select a point \textbf{A} that is vertically aligned with the center of the three V atoms in the Cs plane and simulate the nearby local field distribution.
We designate point \textbf{A} as the center of the $xy$-plane and simulate the local magnetic field distribution in three different directions: $B_{z}, B_{x}, B_{y}$, as shown in Figure.\ref{P-5} (b-d), respectively. The local magnetic field along $z$ direction ($B_{z}$) exhibits the largest magnitude. Furthermore, compared to the local magnetic field along $x$ direction ($B_{x}$), the local magnetic field in others directions are more localized around {\bf A} point.

According to the analysis above, when these anomalous phonons are excited, the phonon moment of V atoms will induce a local magnetic field in Cs plane.
In the $\mu$SR experiment, muons decayed into positrons. These positrons carry energy and specific momentum will scatter with phonons, potentially exciting anomalous phonons. This phenomenon is akin to the excitation of chiral phonon by the photons \cite{Cheng2020-Nano,Nova2017-NaturePhys} and also occurs in the AHE.
Furthermore, we analyze the magnitude of magnetic field intensity and magnetic moment induced by this mode. We estimate that the magnetic field intensity approximately 10$^{-5}$ T at point \textbf{A}, which falls in the range of magnitude in order detected by the $\mu$SR experiment. Meanwhile, the magnetic moment approaches 10$^{-4} \mu_{B}$ per V atom. This value is comparable with the magnitude of the magnetic moment reported in antiferromagnetic materials with considerable AHE \cite{Nakatsuji2015-Nature}.
For a detailed discussion, please refer to the {\it Supplemental Materials} Sec.S6.
Thus, the phonon magnetic moment suggests an explanation on the origin of local magnetic field observed in the $\mu$SR and AHE experiments, and it is worth verifying experimentally.


\emph{Discussions.-}
We have extended the theory of chiral phonon to the anomalous phonon in kagome metals, the global chiral phonon in noncentrosymmetric crystal structures is replaced by the local anomalous phonon in centrosymmetric crystal structures, just as the global magnetization in ferromagnet is replaced by sublattice magnetization in antiferromagnets.
In the CDW phase, we do not observe anomalous acoustic phonons for V atoms; instead, we predict that the significant experimental characteristic will be anomalous optical phonons.
The possible experimental characteristics should be either the Raman spectroscopy or the infrared circular dichroism spectra. As we have shown in the proceeding section, for the former, the energy range of anomalous phonons of V atom fall in 5 THz to 9 THz, and for the latter, the presence of the local phonon magnetic moment will demonstrate in infrared circular dichroism. We anticipate further experiments could verify the prediction of the present theory.

There are two additional points that should be mentioned.
{\bf (1)} The vibrations of V atoms in the neighboring kagome plane contribute to an opposite local field in the Cs plane, resulting in the overall local field being zero along the $x$ and $y$ directions. However, if the CDW phase adopts a $2\times2\times2$ configuration with $\pi$-phase shift rather than the $2\times2\times1$ configuration mentioned above, the entire local field in the Cs plane along the $x$ and $y$ directions would be nonzero.
For further details, please refer to the {\it Supplemental Materials} Sec.S7.
{\bf (2)} Because of the time-reversal symmetry, at {\bf +q} and {\bf -q} points, the vibration directions are opposite for the same atom, resulting in a zero the macroscopical phonon moment. Nevertheless, in the $\mu$SR experiment, muons decay into positrons, creating an effective scattering cross section of positrons and leading to a nonequilibrium state of the system. Consequently, it results in a different number of phonons at $\pm${\bf q} points and a nonzero macroscopical phonon magnetic moment in real space, which further causes the TRSB.
Please see the {\it Supplemental Materials} Sec.S8 for detailed discussion.

Yu {\it et al.}  \cite{Yu2021-PRB} and other authors  \cite{Yang2020-SciAdv, Yin2021-CPL} have reported giant AHE in KV$_3$Sb$_5$ and RbV$_3$Sb$_5$, in their CDW phase, which has sparked significant interest. The absence of local magnetic moment \cite{Yu2021-PRB,Kenney2021-JPCM} in these kagome metals ruled out the skew scattering mechanism, while the intrinsic Berry phase alone is insufficient to account for the observed anomalous conductivity in experiments. This the necessity of a new mechanism.
The presence of anomalous phonons and their associated phonon magnetic moment could provide an explanation for the microscopic origin of the AHE and the TRSB observed in CsV$_3$Sb$_5$ and KV$_3$Sb$_5$. As demonstrated above, the polarized motions of V ions in the CDW phase may contribute orbital magnetic moment of phonons, resulting in extraordinary scattering to the Hall transport.

Moreover, as we address in the beginning, the debate surrounding the microscopic origin of the TRSB could be spontaneously resolved within the framework of present anomalous phonon and phonon magnetic moment. The orbital magnetic moment arising from anomalous phonons contributes to the local magnetic field, thereby breaking time-reversal symmetry.
Hence, the concept of anomalous phonons and phonon magnetic moments could naturally explain microscopic origin of the puzzled AHE and the TRSB effects in the CDW phase of the kagome metals AV$_3$Sb$_5$, elucidate other hidden magnetic properties in distorted materials.

\emph{Summary.-}
We perform the first-principles calculations to investigate the atomic vibrations of the pristine and the CDW phases of CsV$_3$Sb$_5$. Both phases maintain spatial inversion symmetry, thus chiral phonon do not emerge, but anomalous phonons are present in certain high-symmetry paths.
In the CDW phase, the lattice distortion of kagome plane induces polarizations of V atoms, with the magnitude of polarization dependent on the degree of distortion. These polarizations manifest as elliptical vibrations of V atoms within kagome plane, leading to the emergence of orbital magnetic moments and contributing to the local magnetic field. The presence of orbital magnetic moments in phonons offers a novel explanation for observations in the $\mu$SR experiment, even the AHE in CsV$_3$Sb$_5$, and deserved to further investigate in experimental study.

\

The authors thank the supports from the NSFC of China under Grant Nos. 11974354,
Program of Chinese Academy of Sciences, Science Challenge Project No. TZ2016001.
Numerical calculations were partly performed at the Center for Computational Science of CASHIPS,
the ScGrid of the Supercomputing Center, the Computer Network Information Center of CAS,
the CSRC computing facility, and Hefei Advanced Computing Center.

\bibliographystyle{apsrev4-2}
\bibliography{chiral}

\end{document}